# RESEARCH

# Visualization and Correction of Automated Segmentation, Tracking and Lineaging from 5-D Stem Cell Image Sequences

Eric Wait[1], Mark Winter[1], Chris Bjornsson[2], Erzsebet Kokovay[3], Yue Wang[2], Susan Goderie[2], Sally Temple[2] and Andrew Cohen[1*]

**Abstract**

**Background:** Neural stem cells are motile and proliferative cells that undergo mitosis, dividing to produce daughter cells and ultimately producing differentiated neurons and glia. Understanding the mechanisms controlling neural stem cell migration and proliferation will play a key role in the emerging field of regenerative medicine and in cancer therapeutics. Neural stem cell studies *in vitro* are maturing; however, the cell's environment may play a pivotal role in determining cell fate. Thus cells must be observed in their native niche. Visualizing a three dimensional image of an intact tissue sample for quantitative and qualitative analysis is non-trivial. It becomes more difficult when the tissue sample is large and imaged at sub-cellular resolution or when imaged over time.

**Results:** We present an application that enables the quantitative analysis of multichannel 5-D $(x, y, z, t, channel)$ and large montage confocal fluorescence microscopy images. The image sequences show stem cells together with blood vessels, enabling quantification of the dynamic behaviors of stem cells in relation to their vascular niche, with applications in developmental and cancer biology. Our application automatically segments, tracks, and lineages the image sequence data and then allows the user to view and edit the results of automated algorithms in a stereoscopic 3-D window while simultaneously viewing the stem cell lineage tree in a 2-D window. Using the GPU to store and render the image sequence data enables a hybrid computational approach. An inference-based approach utilizing user-provided edits to automatically correct related mistakes executes interactively on the system CPU while the GPU handles 3-D visualization tasks.

**Conclusions:** By exploiting commodity computer gaming hardware, we have developed an application that can be run in the laboratory to facilitate rapid iteration through biological experiments. There is a pressing need for visualization and analysis tools for 5-D live cell image data. We combine accurate unsupervised processes with an intuitive visualization of the results. Our validation interface allows for each data set to be corrected to 100% accuracy, ensuring that downstream data analysis is accurate and verifiable. Our tool is the first to combine all of these aspects, leveraging the synergies obtained by utilizing validation information from stereo visualization to improve the low level image processing tasks.

**Keywords:** volumetric texture rendering; 3-D display; stereoscopic 3-D; stem cell; time lapse; lineaging; validation and correction; confocal microscopy; CUDA; image montage reconstruction

## Background

Neural stem cells (NSCs) are motile and proliferative cells that undergo mitosis, dividing to produce daughter cells and ultimately producing differentiated neurons and glia. Understanding the mechanisms controlling NSC migration and proliferation will play a key role in the emerging field of regenerative medicine and in cancer therapeutics. All of the cells in a clone are genetic copies of the original stem cell. Image-based analysis of static 3-D images demonstrated the important relationship between neural stem cells and blood vessels, and the propensity of both adult and embryonic NSCs to seek out and maintain distinct spatial relationships with respect to vasculature known as their vascular niche [1–3]. Confocal and multiphoton micro-

[*]Correspondence: acohen@coe.drexel.edu
[1]Drexel University, 19104 Philadelphia, US
Full list of author information is available at the end of the article scopes that contain integrated incubation systems are able to image live NSCs together with blood vessels in intact tissue slices, with 5-D image stacks $(x, y, z, t, \lambda)$ captured at intervals (*e.g.* 20 min.) over a period of 16–20 hours. Here, $\lambda$ represents spectral information from a fluorescent label. By labeling the blood vessels and the NSCs with different fluorescent markers, these microscopes are able to capture image sequence data that shows the dynamic behaviors of migrating proliferating NSCs while simultaneously capturing the relationship to other structures including blood vessels. We have developed an application that for the first time enables the use of time-lapse microscopy data to quantify the dynamic relationship between clones of mammalian NSCs and their niche in intact tissue containing vasculature and live proliferating cells.

The analysis of clones of migrating proliferating NSCs starts with *segmentation*, the delineation of individual cells in each image frame. *Tracking* then establishes temporal correspondences between segmentation results. Finally, *lineaging* establishes parent-daughter relationships across mitotic events. The analysis of stem cell clonal dynamics to date has consisted primarily of extracting and analyzing a *lineage tree* generated from cultured cells. A lineage tree is a graphical representation showing each cells' division time and the offspring it produces. Each daughter cell is a genetic copy of its parent cell, a lineage tree is often referred to as a *clone* of stem cells. Lineages also indicate the population dynamics of clones of stem cells, showing the lifespan and parentage of each cell in the clone, as well as indicating the phenotype of differentiated progeny. These trees summarize patterns of division (*i.e.* symmetric or asymmetric, cell cycle time, number of divisions, *etc.*) and differentiation in a single view. The lineage tree is a key tool in the analysis of stem cell clonal dynamics. In addition to tree level features, we can analyze cellular properties such as motion and morphology using tools such as Algorithmic Information Theoretic Prediction and Discovery (AITPD) [4, 5]. AITPD analyzes the patterns of cellular dynamic behavior for individual cells established by segmentation, tracking and lineaging. It can accurately predict development potential for individual NSCs. Previously, we have shown that software in conjunction with AITPD enables the search for behavioral markers of different functional subtypes as well as the potential discovery of molecular mechanisms controlling stem cell proliferation [6]. There is a pressing need for new approaches to visualize and validate 5-D image sequence data of proliferating mammalian cells to enable quantitative analysis of the mechanisms controlling cellular proliferation and differentiation.

While it is possible to analyze the dynamic behaviors of stem cells in a manner that is robust to segmentation errors, **any** *errors in tracking or lineaging are likely to corrupt all subsequent analysis.* For *in vitro* phase contrast time lapse image sequence data (2-D) we recently developed a software tool called LEVER that allows a biologist to run automated segmentation, tracking, and lineaging on image sequence data in the laboratory [6]. LEVER displays the lineage tree in one window, while the image sequence data with segmentation and tracking results overlaid are displayed in a second window. Navigation and editing can be done on either window. The interface is designed so that users are able to easily identify and quickly correct any mistakes in the automated image analysis. We have found in this work and in previous studies that the vast majority of errors in tracking and lineaging are the result of segmentation errors [5, 6]. These errors happen often when the number of cells in a given area have been incorrectly estimated. LEVER uses an inference-based learning approach, which propagates user-supplied corrections forward to reduce errors on later frames. Here we present an application called LEVER 3-D which displays image data in full stereoscopic 3-D and includes a utility to reconstruct 3-D image montages with the *intent to be run in the biology lab*. LEVER 3-D uses commodity gaming hardware to implement a high-performance interactive system for validating and correcting the automated image analysis results for 5-D stem cell data. This will allow the biologist to better understand stem cell dynamics and regulation within the neural stem cell microenvironment.

## Methods

A total of 18 5-D image sequences showing adult mouse neural stem cells, ependymal cells, and vasculature were analyzed. The stem cells were imaged within a 3D wholemount explant culture of the subventricular zone (SVZ) of the brain. Each 5-D voxel location is specified as $(x, y, z, t, \lambda)$ where $\lambda$ refers to a multichannel fluorescence signal, one channel imaging NSCs, the second channel containing ependymal cells and vasculature. The movies were captured on two microscopes at two different laboratories. SVZ wholemounts were dissected under a dissection microscope as described previously [7, 8] from transgenic mice that express green fluorescent or tomato red fluorescent protein in neural stem cells (FVB/N-Tg(GFAPGFP) 14Mes/J, the Jackson Laboratory; Ascl1$^{\text{tm1.1(cre/ERT2)}}$; B6.Cg-Gt(ROSA)$^{\text{26Sortm9(CAG-tdTomato)}}$, NSCI). Briefly, the brain was removed and halved and the cortex was peeled back to reveal the SVZ. A scalpel was used to make a 2-4 mm cut on the striatal side of the SVZ and watchmaker forceps were used to clip off the SVZ at the anterior and posterior sides and carefully transferred into phosphate buffered saline containing 5 $\mu$g/ml Alexa Fluor 647 conjugated Isolectin GS-IB4 (Life Technologies) on ice for 20 minutes to label the ependyma and blood vessels. SVZ wholemounts were transferred SVZ side down to a glass bottomed culture dish (MatTek Corporation) and immobilized by covering with cold (4°C) growth factor free Matrigel (BD Biosciences) and immediately transferred to an incubator set at 37°C with 5% CO2 for 20 minutes to allow the Matrigel to solidify. Freshly made slice culture medium was added to the culture dish and the dish was placed on a Zeiss LSM780 confocal microscope equipped with an environmental enclosure set at 37°C and 5% $CO_2$. The pinhole was set at 2 AU and Z stacks (25 1 $\mu$m steps) were collected every 20 minutes using a 20X objective for 16 hours. Spatial resolution was 512×512, at a pixel spacing of 0.8 $\mu$m for a total of 1.3 GBytes of image data at 32 bits per voxel (BPV). At this resolution, the entire image sequence can be downloaded to the video RAM on a 1.5 GB card, with room left for the frame buffer and back buffer. Images were also captured on a Zeiss LSM510 confocal microscope at a spatial resolution of 1024×1024 at a pixel spacing of 0.3 $\mu$m for up to 20 hours, resulting in as much as 5 GBytes of image data per sequence. These larger sequences require the image sequence data to be buffered in system memory, a task that is handled automatically by the display driver. Interestingly, less proliferation was observed in image sequences captured at the higher spatiotemporal resolution. Once the 5-D image sequence data has been captured it is imported directly from the microscope output file using the open source Bioformats application.

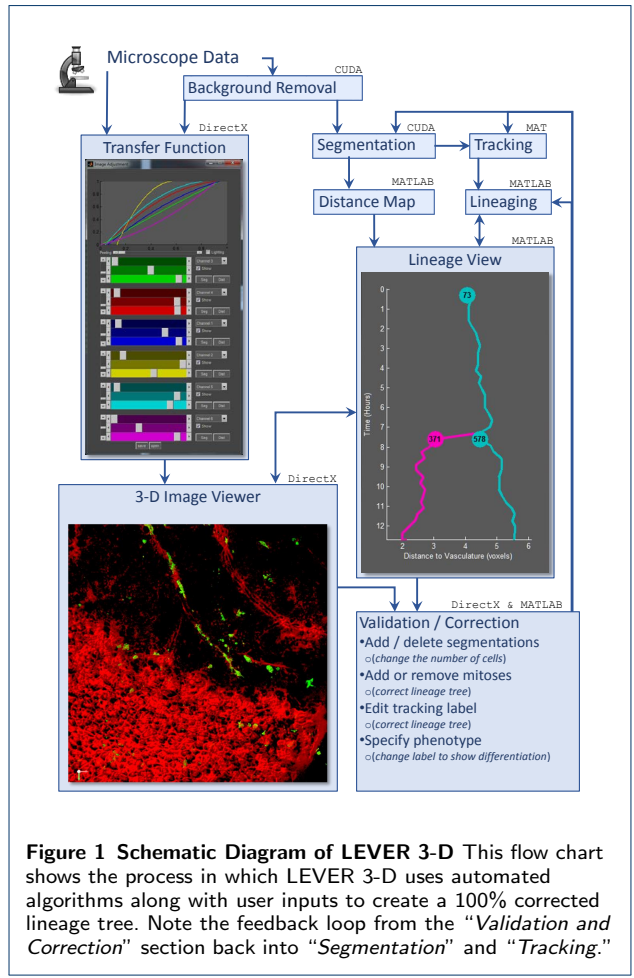

**Figure 1 Schematic Diagram of LEVER 3-D** This flow chart shows the process in which LEVER 3-D uses automated algorithms along with user inputs to create a 100% corrected lineage tree. Note the feedback loop from the "*Validation and Correction*" section back into "*Segmentation*" and "*Tracking*."

Processing of the 5-D image sequence data begins by using the open-source BioFormats [9] tool to convert the native microscope data (Zeiss LSM file) into intensity valued tiff images. The use of BioFormats enables LEVER 3D to work not only with the Zeiss specific file formats, but with file formats used by most of the major microscope manufacturers. In addition to the image data, BioFormats extracts the imaging settings including the spatiotemporal resolution used to account for image anisotropy, providing scaling for the tracking and distance-based calculations.

Figure 1 illustrates the flow of data and processing steps used to go from the raw input image sequence data to a fully validated and corrected clonal tracking and lineaging. Figure 1 also illustrates the main software components used including CUDA for efficient image processing from C++, MATLAB for 2-D visualization of the lineage tree and data analysis and export, and Direct 3D for 3-D rendering and visualization. Each of these steps is described in more detail throughout the remainder of this section.

Background Noise Removal

Confocal and multiphoton fluorescence microscopy of live stem cells is always battling between capturing enough signal and the disturbance of the specimen. Imaging is done in a manner that maximizes the signal-to-noise ratio (SNR) of the image data while being minimally invasive. Using less excitation energy causes less phototoxicity and disturbance to sensitive proliferating cells. This means that in practice, SNR ends up being quite low. One way to improve the analysis of this large volume of challenging image sequence data is to apply pre-processing techniques that model the underlying dynamic data and noise processes. Here we use two different background noise removal techniques for the stem cell and the vasculature channel to better match the characteristics of the objects being imaged. These background noise removal algorithms provide the benefit both of removing noise and of providing adaptive contrast enhancement. This simplifies and improves the performance of the subsequent visualization transform as well as the cell and vessel segmentation algorithms.

For the stem cell channel, we adapt the background noise removal technique described by Michel *et al.* [10] that models the noise as a slow varying low-frequency background noise signal combined with a random (shot) noise process. Our approach differs slightly in that rather than using extreme value theory and peaks over thresholds approach to detecting fluorescent particles, we use a segmentation approach based on mathematical morphology combined with an adaptive Otsu transform [11] on the filtered image as described in the "*Segmentation*" section. Given the observed image ($\tilde{L}$) that we model as a combination of low-frequency background ($B$), random shot noise ($R$) and the original (denoised) image ($\hat{L}$) that we wish to recover,

$$\tilde{L} = B + R + \hat{L}. \tag{1}$$

The low-frequency background contribution ($B$) is estimated using a low pass (Gaussian) filter. After subtracting the estimated background component from the observed image, the high frequency shot noise is removed using a median filter to produce the final denoised image used in the visualization transfer functions and the segmentation algorithm.

The above denoising approach works well on the stem cell channel where the foreground voxels (pixels with a third dimension) are found in relatively small high frequency regions corresponding to cells. In the vasculature channel, foreground voxels can constitute large portions of the image corresponding to dense regions of blood vessels. For denoising the vasculature channel we have therefore adopted a different denoising approach using Markov random fields [12] and a global estimate of noise variance rather than the local background estimate used in the stem cell channel. This is an iterative technique that first estimates the noise variance $\tilde{\sigma}$ for the original image $I^0$ by convolving with a Laplacian operator. This noise variance is used as the stopping condition for our iterative de-noising where we keep the $n^{\text{th}}$ convolved image,

$$\|I^{n+1} - I^0\| \leq \tilde{\sigma}. \tag{2}$$

We compute the minimum value $\Delta$ to be used as a step size in adjusting the intensity at each voxel, $\Delta = \min_{q,r \in i, q \neq r} |v_q - v_r|$ where $v_i$ is the finite set of voxel values in the image. Each iteration adjusts (subtracts or adds depending on the sgn function in equation 3) the intensity of every voxel by a $\Delta$ depending on the gradient of the neighborhood, as defined as:

$$\begin{aligned}I^{n+1}_{i,j,k} = I^n_{i,j,k} + \Delta \times \\ \text{sgn}\bigg[ \text{sgn}(I^n_{i-1,j,k} - I^n_{i,j,k}) + \text{sgn}(I^n_{i,j,k} - I^n_{i+1,j,k}) \\ + \text{sgn}(I^n_{i,j-1,k} - I^n_{i,j,k}) + \text{sgn}(I^n_{i,j,k} - I^n_{i,j+1,k}) \\ + \text{sgn}(I^n_{i,j,k-1} - I^n_{i,j,k}) + \text{sgn}(I^n_{i,j,k} - I^n_{i,j,k+1}) \bigg]. \end{aligned} \tag{3}$$

The background denoising algorithms for both the vasculature and stem cell channels simplify the subsequent segmentation algorithm and also the transfer function used in adjusting voxel intensity and transparency values for visualization.

Registration

In previous studies, subsections of the subventricular zone (SVZ) have been targeted for inspection [13]. Subsections were necessary due to the field of view being small given high magnification. We would rather be able to inspect the structure of the entire SVZ at the highest resolution possible. Having a sub-cellular resolution means that we can compare different populations with a higher level of precision. However, subsections may not correspond exactly between experiments. Our solution to this problem is to break the SVZ into a mosaic of high resolution overlapping subsections. These images are then reconstructed into an ultra high resolution large volume image of the entire structure.

With modern microscopes, stage position is stored in the metadata of each image capture. Knowing the approximate stage position of the subsections with respect to one another, we can overlay each into the entire structure. However, this result leaves much to be desired. The initial reconstruction can be viewed as the green lines in Figure 2. During the imaging process, the specimen can shift relative to the stage, making stage position insufficient for registration. This can be caused by vibration of the stage, mechanical drift, dehydration or settling of the tissue slice, as wells as removal of the slide between images. These position inaccuracies can be quickly overcome by using the overlapping image regions to find the relative offsets (register) the mosaic of images.

The complexity of the registration problem can be reduced by exploiting the fact that the images consist of a single time-point and that each image in the montage is oriented orthogonal to the imaging stage. A montage with a large number of subsections, approaching 100, implies that the specimen has to be static in time. Imaging one subsection is time consuming and by the time another column or row is started, the overlap sections would have changed too much to be reconstructed. The second assumption, enforced by the microscope mechanisms, is that the subsections are all at right angles to one another in a checkerboard fashion. Lastly, the volume will only deform generally in one direction, which is consistent with settling and dehydration where the volume will deform in the direction of gravity. This alleviates the need for transformations such as rotation, shearing, and deformation when registering. Based on these assumptions we can formulate an effective and accurate translational registration algorithm.

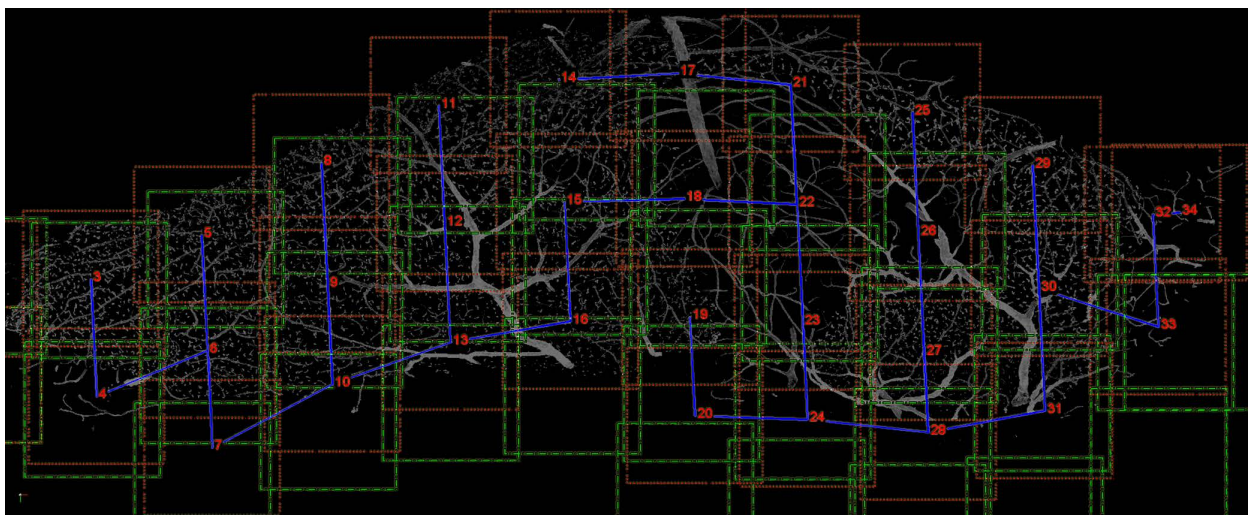

**Figure 2 Reconstruction of an Entire SVZ** The image is the result of registering 34 subsections of a mouse subventricular zone. Registration using only microscope stage position data is indicated with a green dashed line. The blue solid lines represent a max spanning tree indicating which edge of the subsection was registered, *e.g.* subsection 22 was registered to 18, 21, and 23, where subsection 11 was only registered to 12. The red dashed lines indicate the final position of each subsection after registration. Registration happens in the $z$ direction as well, not shown here.

With multichannel fluorescent microscopy, there typically exists a channel with unique semi-sparse structures that will span image subregions. The SVZ's vasculature channel is an ideal example of such a structure. The overlap regions of the selected channel are shifted relative to each other and evaluated to determine how well they fit together. We use the *normalized covariance* between the two overlapping volumes, normcov in Equation 4, to quantify how similar they are. We evaluate a windowed area around the stage position data and choose the new position that maximizes our metric. The benefits to this method are that it is robust to variation in imaging parameters such as intensity (by subtracting the mean $\mu$) and contrast (by normalizing over the variance $\sigma$), as well as being object agnostic. This technique also works directly on the images rather than requiring preprocessing to determine a set of feature points that are used in more complex registration schemes.

$$\text{normcov}(A, B) = \frac{\sum (A - \mu_A) \times (B - \mu_B)}{\sigma_A \times \sigma_B} \quad (4)$$

Each overlapping subsection is evaluated and stored in a graph structure. The nodes represent an offset from the original stage position. Each edge is the normalized covariance measure at the given offset. We then drop the lowest scoring edges until we are left with a max spanning tree, represented by the blue line in Figure 2. This allows us to anchor a single image and follow this max spanning tree to assign a delta value relative to the change in the previous node in the graph. In other words, a node is chosen to be stationary (position based solely on stage position data). The delta for each node connected to this root node is based only on the registration position change. Each subsequent delta on a path of edges is calculated from the cumulative delta from the root and local registration delta. The final positions can be viewed as the red lines in Figure 2. Once every subimage's delta has been calculated, a final reconstructed image can be created. Given that each channel of a particular subimage has been taken at the same time, we can register every channel using just one delta value. Figure 3 shows a fully registered SVZ containing five channels. Supplemental video labeled *SVZ.mp4* shows the same volume rotating and zooming into a region of interest. These images are then exported as full sized tiff files along with updated metadata files which contain the new dimensions. The image in Figure 3 has dimensions of 10,173 pixels in $x$, 3,858 in $y$, 74 in $z$, and five channels. These newly exported images are now ready to be processed just like any other images received from the microscope.

### Segmentation

In previous applications involving stem cell segmentation and tracking in phase contrast 2-D image sequences [5, 6, 14] we have found that the most significant challenge to segmenting stem cells is to identify the correct number of cells in each connected component of foreground pixels. On a given frame, the numbers of cells in a given area maybe ambiguous to even a domain expert. This ambiguity is typically the result of cells touching. This typically occurs immediately following a mitosis event or when there is a high density of cells. The problem of estimating the number of cells in any connected component of foreground voxels is easier in 3-D due to the discriminative nature of the extra spatial channel. The output of the denoising algorithm removes data that is not directly related to foreground voxels, and is especially useful at preserving the gradient boundaries between cells. This improves the accuracy of the subsequent segmentation algorithm.

Our segmentation algorithm begins by applying an adaptive thresholding, to all channels, using a CUDA Otsu filter [11]. This results in a binary image of foreground and background voxels. A morphological closing operator using a binary ball structuring element is applied to remove any erroneous holes in the structures. The stem cell channel is next processed with a connected component image filter and any connected components less than $19\mu m^3$ in volume are discarded. This value can be set by the user prior to running the segmentation algorithm. Finally, the convex hull of the foreground voxels for each cell is computed using the open-source QHULLS package [15], creating facet and vertex lists for each cell. The convex hulls generated by QHULLS are then loaded into Direct 3D vertex and index buffers.

The vasculature channel is processed in a similar manner. Following the adaptive thresholding, a distance map is computed using a distance map filter. This provides the distance from each voxel to the nearest foreground voxel, and is used in the subsequent analysis to quantify the distance between each cell and the nearest vessel. The results of the stem cell segmentation are next passed to the tracking algorithm to establish temporal correspondences between segmentation results and assign tracking IDs to each cell.

### Tracking

Once the cells have been segmented we use an approach called Multitemporal Association Tracking (MAT) [6, 16] to establish temporal correspondences among segmentation results. MAT is a graph-based tracking approach, which, for each frame, evaluates a multi-temporal cost function that approximates the Bayesian *a posteriori* association probability between the current set of tracks and all feasible track extensions out to a fixed window size $W$. In multiple hypothesis tracking, this data association problem is typically solved using bipartite or multidimensional assignment, which is NP-hard and requires explicit modeling of imaging specific conditions including occlusions, missed and extraneous detections. MAT instead uses a minimum spanning tree approach to solve the data association problem, relying solely upon typical cell dynamic behavior of smooth motion and independent of imaging conditions. In addition to the current results tracking stem cells in 3-D, MAT has achieved excellent results for hundreds of *in vitro* (2-D) image sequences of mouse adult and embryonic neural stem cells, as well as hematopoietic stem cells and rat retinal progenitor cells [6] and has been applied to tracking high density organelle transport along the axon [16]. All of the stem cell movies tracked by MAT in 2-D and 3-D were processed with the same implementation.

In order to extend tracks from frame $t$ to $t+1$, we denote a partially constructed track terminating at the $i^{\text{th}}$ detection in frame $t$ as $\tau_i^t$, and denote the set of all feasible extensions passing through the $j^{\text{th}}$ detection in frame $t+1$ as $\rho_i^{t+1}$. The cost of edge $c_{ij}$ in the tracking graph is assigned the minimum cost of extending partial track $\tau_i^t$ through the $j^{\text{th}}$ detection in the next frame $c_{ij} = \min_{t+1}\{C(\tau_i^t, \rho)\}$. Edges $c_{ij}^\star$ satisfying $c_{ij}^\star \leq c_{in}$ and $c_{ij}^\star \leq c_{mj}$ are called matching edges. We extend each track $\tau_j^t$ along its matching edge $c_{ij}^\star$. For any detections in $t+1$ without a matched incoming edge $c_{mj}^\star$, we initialize a new track. Occlusions, where one cell is visually obscuring another, are handled by allowing tracks that are not extended to be considered in subsequent frames.

For tracking stem cells in the 5-D image sequences we used the same cost function used previously for tracking 2-D phase-contrast imaged NSCs [6], with the sole modification of using a Z value in the connected-component distance. We define the connected component distance between two detections as

$$d_{\text{CC}}(\alpha, \beta) = \min_{\rho_\alpha \in \alpha, \rho_\beta \in \beta} \{\|\rho_\alpha - \rho_\beta\|\}, \quad (5)$$

where $\rho_\alpha$, $\rho_\beta$ are the scaled-voxel coordinates corresponding to the foreground-voxels of the segmentation detections and, respectively. We also define a detection size distance to preserve homogeneous sizes along a given track,

$$d_{\text{size}}(\alpha, \beta) = \frac{\max\{|\alpha|, |\beta|\} - \min\{|\alpha|, |\beta|\}}{\max\{|\alpha|, |\beta|\}}, \quad (6)$$

where $|\alpha|$ is the number of pixels in the foreground connected-component of detection $\alpha$. For a given path $C(\tau, \rho)$ extension we calculate the cost of the extended track $(\tau, \rho)$, as a weighted sum of the local connected component distances along the detections of $(\tau, \rho)$,

$$C(\tau, \rho) = (W - |\rho| + 1) \times \left( \sum_{i=-1}^{|\rho|-1} w_i \Big( d_{\text{CC}}(\rho_i, \rho_{i+1} + d_{\text{size}}(\rho_i, \rho_{i+1})) \Big) \right), \quad (7)$$

with $\rho_i$ indicating the $i^{\text{th}}$ detection on path $(\tau, \rho)$. By convention, if $i \leq 0$ we use $\rho_i = \tau_{\tau_{t+i}}$, which allows evaluation of the cost over the full-extended path. This cost reflects the expected behavior of neural stem cells, namely, the cells should not move far (small connected-component distance) and their size shouldn't vary greatly in adjacent frames. The multiplicative term discourages shorter tracks which would otherwise have lower cost due to fewer terms. We use a window size $W = 4$ and weights $w = [1\,3\,1\,1\,1]$ for $i = -1\ldots 3$. This cost function has proven effective in tracking both 2-D and 3-D cells as shown in the current and previous applications.

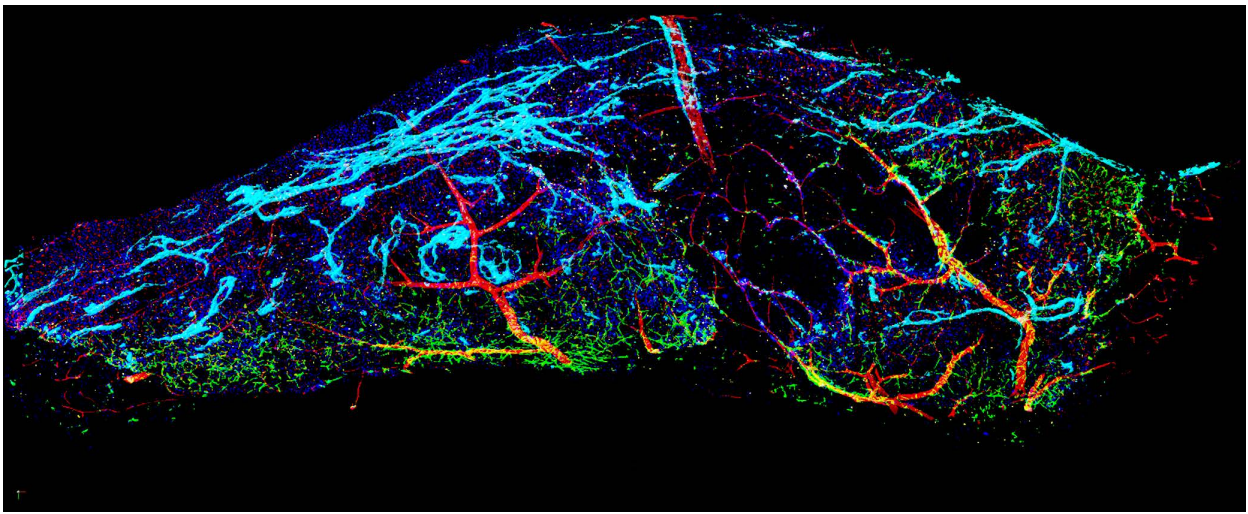

**Figure 3 Fully Registered 3-D Montage with 5 Channels** This image has been reconstructed and rendered using the 3-D view window with adjustments made in the transfer function interface in Figure 5. The channels are: blood vessels (red), cell nuclei (dark blue), neural stem cells and astrocytes (green), oligodendrocytes (yellow), and migrating neuroblasts (cyan).

### Lineaging

Lineaging identifies parent-daughter relationships among the proliferating cells. Our lineaging approach uses the minimum cost path extensions discovered by the MAT algorithm and stored in a sparse graph structure during tracking, and is the same algorithm we have previously used for 2-D lineaging [5, 6, 14]. Cells that constitute viable tracks that appear after the first image frame are assigned parents based on these tracking results. Given a track $\tau_{\text{newborn}}$, we identify its parent track $\tau^*_{\text{parent}}$ as

$$\tau^*_{\text{parent}} = \arg\min \Big\{ C(\tau_{\text{parent}}, \tau_{\text{newborn}}) \Big\}. \quad (8)$$

Once this has completed for all tracks in the image sequence, the largest (most nodes) lineage tree is presented to the user. This tree is the first one to be displayed with the expectation that it represents either the most interesting biological lineage or the lineage that needs the greatest number of user edits.

Traditional lineage trees communicate informative data such as cell cycle time, number of progeny, apoptosis, symmetry of subtrees, *etc*. When studying cells in their niche, we are often interested in spatial relationships between objects. In Figure 4 we are interested in the distance between stem cells and their nutrient source. Here we perturb the vertical lines on the lineage tree in the direction of the x-axis to show the distance between a particular stem cell and its nearest vasculature voxel. Here you can see that one daughter cell has split away towards a vascular structure while the other cell keeps its distance. We also represent the angle in which the cells cleave during mitosis by a plane. This cleavage plane allows for qualitative inspection of how cells divide as well as an additional feature for quantitative analysis.

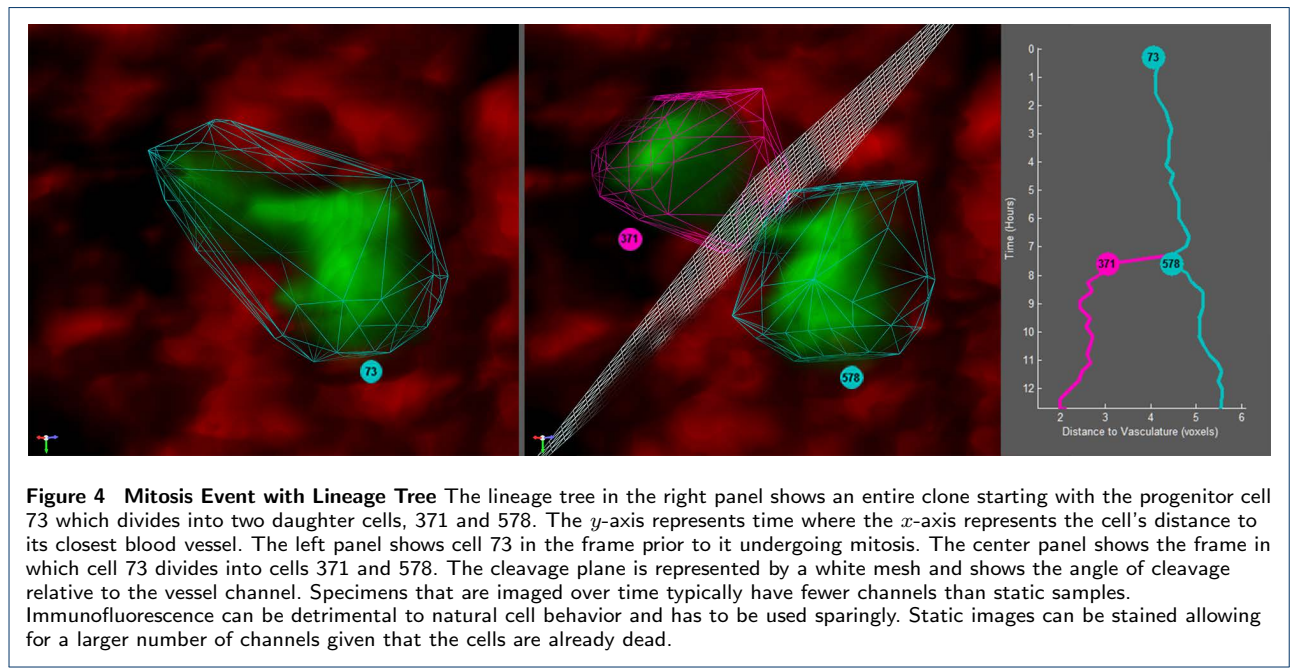

**Figure 4 Mitosis Event with Lineage Tree** The lineage tree in the right panel shows an entire clone starting with the progenitor cell 73 which divides into two daughter cells, 371 and 578. The $y$-axis represents time where the $x$-axis represents the cell's distance to its closest blood vessel. The left panel shows cell 73 in the frame prior to it undergoing mitosis. The center panel shows the frame in which cell 73 divides into cells 371 and 578. The cleavage plane is represented by a white mesh and shows the angle of cleavage relative to the vessel channel. Specimens that are imaged over time typically have fewer channels than static samples. Immunofluorescence can be detrimental to natural cell behavior and has to be used sparingly. Static images can be stained allowing for a larger number of channels given that the cells are already dead.

User Interface
Automated analysis of image sequence data showing proliferating cells will inevitably make mistakes. In addition to low SNR, these image sequences often contain visual ambiguities, *e.g.* due to cells dividing, entering or exiting the imaging frame *etc.*, where it may be impossible even for a human domain expert to correctly identify the number of cells in a connected component of foreground voxels from a single time point. These segmentation errors cause tracking and lineaging errors which can then corrupt the ultimate analysis. Once all of the automated image analysis algorithms have completed, the data needs to be displayed to the user in such a way that it enables errors to be easily identified and quickly corrected. The approach adopted here is to display the volumetric image sequence data with segmentation and tracking results overlaid in a Direct 3D window and the lineage tree in a second MATLAB interactive figure window. The Direct 3D and MATLAB components run in a single memory process, launched from MATLAB and communicating with shared memory via the MATLAB Mex interface.

Visualization of the volumetric data uses the notion of 3-D textures. The slices of each 3-D texture are projected onto planes which consist of two view aligned triangles. There are $\sqrt{2}$ times the maximum pixel dimension of these planes spanning the depth of the image volume. Direct 3D maps the image data onto the triangles using a custom shader. This shader incorporates parameters derived from the transfer function sliders in Figure 5. This user interface was inspired by the work of Wan *et al.* [17]. Currently there are six unique colors that each channel can be assigned. As Wan *et al.* point out, it is difficult to render unique visually separable colors greater than three. This becomes even more difficult when users are allowed to adjust the alpha channel of the voxels. LEVER 3-D allows any of the colors to be assigned to a channel and its visibility toggled.

The colored sliders for an assigned channel are used to set a polynomial transfer function that will map the original image intensity values to an intensity value that will be colored and displayed in the 3-D window. The darker (top) slider for a given channel will set where the low values of the original image will be floored to zero. The brightest (bottom) slider for a given channel is used to set the threshold in which all larger values are mapped to the maximum value; in this case 255 for an 8-bit image. The middle slider changes the curve of the line between the max and min values. The slider on the left edge is used to set the multiplier on the base alpha value for a given channel, allowing a particular channel to be more or less transparent relative to the others. The base alpha for a given pixel is set to the maximum channel intensity at that voxel after the transfer function has been applied. This slider operates on the range of $[0, 2]$, where the center position equals a multiplier of 1. The last user interface control that pertains to the image data is whether to light the texture or not. The lighting check box turns on and off a global directional light. The image data is lit using the surface normal of each voxel. This normal is approximated directly in the shader by finding the gradient direction for each pixel based on a $3 \times 3 \times 3$ neighborhood. Lighting of this kind gives a more three dimensional feel to the image even when viewed on a two dimensional medium (see Figure 5). Once the images are set to the user's liking, they are able to integrate the segmentation results into the 3-D window.

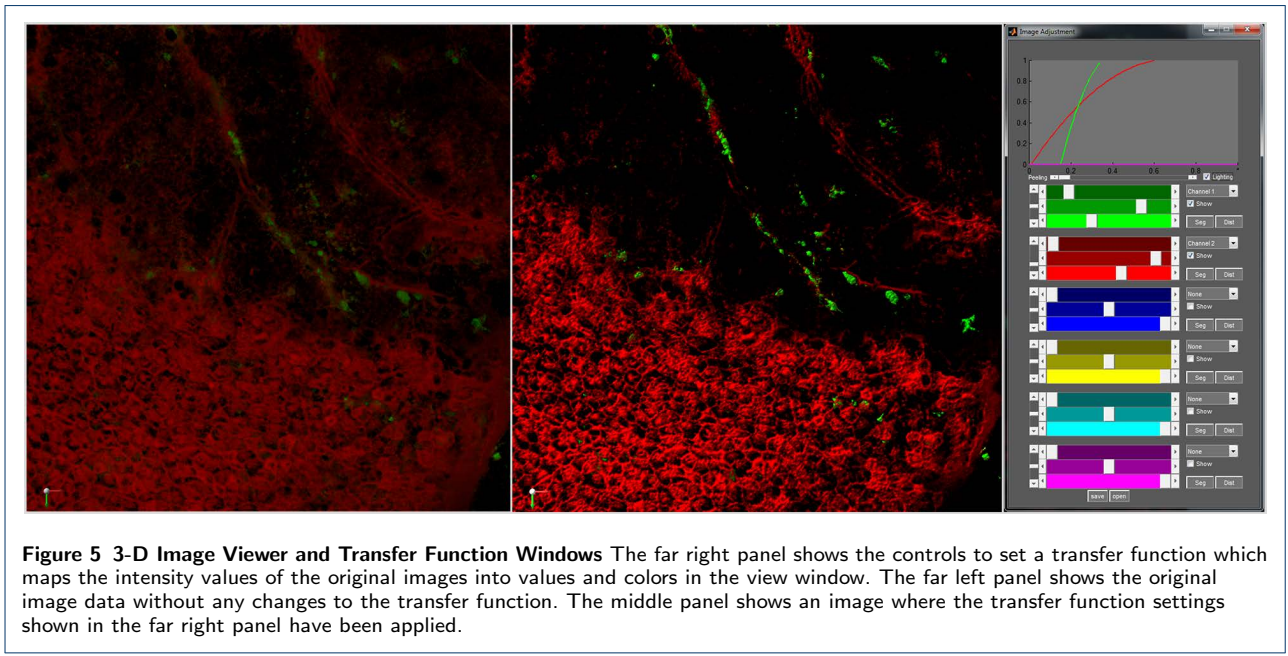

**Figure 5 3-D Image Viewer and Transfer Function Windows** The far right panel shows the controls to set a transfer function which maps the intensity values of the original images into values and colors in the view window. The far left panel shows the original image data without any changes to the transfer function. The middle panel shows an image where the transfer function settings shown in the far right panel have been applied.

Segmentation data is then loaded into video RAM to be overlaid onto the image volume to show tracking and lineaging results. The triangle meshes generated by QHULLS as the convex hull of each cells' voxels by the segmentation process are loaded into Direct 3D index and vertex buffers (triangle lists) and colors are assigned according to tracking and lineaging results. The segmentation buffer is rendered by a second custom shader that can be toggled between off, wire-frame, and solid. The default renderer draws the texture in its entirety and then draws the segmentation results on top of that. This means that the segmentation triangles will always be drawn on top of the image data. This does not show when segmentation is embedded into another structure; however, it renders at high frame rates. Objects obscure one another based solely on their placement in the render loop. Depth peeling gives more visual cues that one object is obscured by another. This is accomplished by layering or "peeling" from each structure in a depth sorted order and interlacing them in the render loop. That will draw the objects that are closer to the view after the further objects, obscuring more distant objects. This can slow the rendering of larger volume data down to non-interactive speeds. To mitigate this problem, a chunked depth peeling has been implemented [17]. There is a slider on the transfer function window (Figure 5) labeled "*Peeling.*". This allows for $[1, N]$ chunks to be peeled, where $N$ is the view aligned voxel count of the volume. This slider can be used to add a level of segmentation integration in the image data as the user's hardware will allow for and still be interactive. The segmentation results are quite often the reason for errors in the tracking and lineaging and are where the majority of the edits take place. This is why it is important to give as many visual queues as possible to show where the segmentation is wrong. When it is wrong, LEVER 3-D allows the user to correct it manually.

Selecting a cell for visualization or editing in the 3-D volume with the use of two dimensional tools (*e.g.* mouse pointer) can be challenging. When the user clicks on the volume, an inverse projection is used to find the intended cell. The inverse projection consists of a ray starting at the view origin passing through the cursor's position on the projection plane and continuing through the volume space. The cell containing the first triangle that is intersected by this ray is then selected. With this selected cell, the user then can remove all other segmentations from the display to leave emphasis on the cell in question. In this view config-uration, the user has the ability to play the sequence and follow the particular cell through the experiment.

Once a cell has been selected, the user then can correct the segmentation or tracking results for the cell. For correcting segmentation results, a cell can be split into $n$ cells or can be deleted. To split a cell, we apply a $k$-means clustering on the foreground voxels of the cell. This is effective because the spherical decision boundaries of the $k$-means algorithm naturally approximate the spherical shape of 3-D NSC's. After a segmentation has been corrected, the tracking automatically reruns. The segmentation edit provided by the user can be "propagated" by inspecting tracking assignments for the original segmentation as potential automated correction candidates until the newly added segmentation establishes its own track. This is the same inference-based approach to learning from user provided edits used in our previous 2-D stem cell lineaging application [6].

The selection of a particular cell also selects the clone to show in the 2-D lineage tree window. The lineage tree is one of the easiest ways of identifying errors in the automated image analysis routines due to predictable qualities such as regularly spaced mitotic events, cells on the lineage tree existing until they reach the end of the sequence or a frame boundary, *etc.* The selection of a cell is translated through a Mex interface to MATLAB allowing the selected clone to be toggled. The shared memory Mex architecture enables all of the segmentation, tracking, and lineaging results to be accessed directly as MATLAB data structures and leverages the implementation from our previous 2-D stem cell lineaging application [6] for lineage tree manipulation and display. The Mex interface allows bidirectional communication between the MATLAB and the Direct 3D user interfaces allowing the two windows to be tightly coupled and enabling a high throughput approach to validating and correcting the automated image analysis algorithms.

The typical work flow of LEVER 3-D would begin with the raw data file from the microscope. The user opens up LEVER 3-D either from Matlab or a compiled version of the program. The benefit of opening LEVER 3-D from Matlab, is that the user has full access to the underlying data structures. The image data is then buffered onto the graphics card and is displayed in the image window. If the current dataset has been processed previously (from a previous session), the segmentation results are rendered in the image window as well as the largest lineage tree in a second window. If the dataset is unprocessed, the user is able to specify a processing method on a particular channel. Now the user is free to explore the image and validate the automated processes using mouse movements. What is viewed in the image window is manipulated by way of the transfer function window. Once the user is satisfied with data and the view settings, LEVER 3-D then can export image, movies, and metrics for external use.

## Results and Discussion

The analysis of the image sequence data proceeds as follows. All timing information is based on a Windows PC with dual Xeon X5570 processors (2.9 GHz), 24 GB of RAM and an Nvidia GTX 430 video card with 2 GB of video RAM. The automated image analysis routines were implemented in C++ using CUDA. Background noise removal, segmentation, tracking, vascular distance and lineaging are run off-line. This step requires from 10 minutes to 2 hours of processing time depending on the quantity of image sequence data. The vast majority of this time is consumed by the background noise removal, a task that is only run once per image sequence and improves the results of the subsequent automatic segmentation algorithm, especially with suppressing image noise between closely adjacent cells improving the ability of the segmentation algorithm to separate nearby cells.

After the automated image analysis routines complete, the 3-D image data with segmentation and tracking results overlaid are shown in the imaging (Direct 3D) window and the lineage tree is shown in a MATLAB figure window connected via MATLAB's Mex interface. The active shutter stereoscopic 3-D visualization glasses improve the visualization of the stem cell data, and especially the relationship between stem cells and vasculature.

Image sequence data displays at 60 frames per second, and manipulation of the 3-D volumetric data is fully interactive even with the 3D stereo vision hardware activated. Navigation can be done on either the 3-D window or on the 2-D lineage window. Clicking in the lineage tree window causes the frame to advance to the selected time point. The time can also be navigated on the Direct 3D window using the mouse wheel, and causes the time indicator on the lineage tree window to update. Users can edit the segmentation and tracking in the imaging window by splitting cells with the mouse or by typing tracking numbers directly onto a cell. The user has only had to correct the automated processes 7% of the time on average. Most of these errors were due to the cells not separating immediately after mitosis. The tracking and lineaging algorithms are executed in response to user provided segmentation edits in order to dynamically update the results and also to correct related segmentation errors in future frames. This process typically requires less than a few seconds to complete, making response to editing operations as well as the 3-D visualization fully interactive.

Once the tracking and lineaging for a clone of stem cells has been corrected, the data can be exported to MATLAB for further analysis. In order to explore the relationship between stem cells and their vascular niche, we used a distance map of the vascular channel. For each stem cell on the clone, we can use this distance map to instantly find the distance between the cell and the nearest blood vessel. In Figure 4 we plot this distance for the three cells on the lineage tree of the selected clone. Cell 73 is at a stable distance to the nearest blood vessel. When the cell divides one of the daughter cells moves into contact with the vessel while the second daughter cell is initially situated farther from the vessel. This is a result of the cleavage plane, formed by the division between the two daughter cells, being oriented acutely toward the vessel. Interestingly, the daughter cell that is closer to the vasculature following division, cell 371, has a different pattern of motion than the daughter cell that is not in contact with vasculature. This may be indicative of a different sub type of stem cell or of the cell seeking to re-establish its location in the vascular niche following division. This is the first time, to our knowledge, that this relationship between a clone of mammalian NSC's and their vascular niche has been visualized and quantified dynamically in live cell and tissue image sequence data. Supplementary Video labeled *Lineage.mp4* shows a segmentation and tracking for this clone simultaneously with the plot of the distance to vasculature for each cell.

A key decision in the design of our application was the use of Direct 3D rather than OpenGL to provide 3-D rendering. In general, scientific visualization applications tend to use OpenGL while gaming applications tend to use Direct 3D. This decision was primarily based on the need to incorporate support for NVidia's 3-D Vision to utilize active shutter stereoscopic glasses into our application. Using Direct 3D rather than OpenGL enables the use of 3-D vision on less expensive NVidia GTX-class gaming cards, as well as on the more expensive Quadro cards. Additionally, automatic driver optimized support for stereo separation is available to Direct 3D applications only [18]. The 3-D vision stereo glasses can be used from OpenGL, but that requires the use of the Quadro card and also requires explicit application support for stereo via quad buffering. Stereoscopic viewing enables a user to quickly identify and easily correct tracking and lineaging errors in an natural and highly interactive manner. Shortcomings of using Direct 3D are discussed in the "*User Interface*" section.

Other applications have been developed for 3-D stem cell lineaging, notably by Murray *et al.*[19]. They developed an approach that does not however, include capabilities for learning from user supplied edits or 3-D visualization. In later work they incorporated a support vector machine to automatically identify segmentation errors [20], although we have found that segmentation errors occur primarily when there is visual ambiguity in the image data that the human eye is unable to resolve using only a single image frame. The current project is an extension of the LEVER application designed for 2-D phase contrast stem cell image sequences which uses a human observer to assist in correcting the visual ambiguity inherent in image sequences of live proliferating cells [6]. Aside from the 3-D rendering, one other difference of the current work is that the segmentation is implemented using CUDA rather than MATLAB. CUDA provides a significant performance improvement over MATLAB, making the 3-D noise and background removal and segmentation algorithms approximately 20 times faster.

In the area of biological image sequence data visualization there are a large number of commercial and open source products, as described in a review paper by Walter *et al.* [21]. One thing that differentiates our work from the described approaches is the tight integration between the automated image analysis algorithms and the Direct 3D visualization. In contrast, most other applications utilize the Visualization Toolkit (VTK) an open source visualization library [22]. The 3-D rendering for our application was initially implemented using VTK, however VTK is compatible with OpenGL but not with Direct 3D. The open source ICY application [23] uses VTK for visualization and provides an extensible user interface for visualizing 2-D and 3-D images and incorporates segmentation and tracking algorithms, as well as editing of results and multiple linked views. Our work differs from ICY in supporting stem cell lineaging and using inference-based learning to propagate user-provided edits. A related application for visualizing multichannel fluorescence microscopy data for biological applications was presented by Wan *et al.* [17]. Their application provided more control over the viewing of the volumetric data including a user controllable 2-D transfer function for setting the rendering properties of the volumetric data. In contrast, the approach presented here uses the automatic image analysis algorithms to set the parameters on the visualization transfer function with the intention that our application will be used for quickly validating and correcting the clonal tracking and lineaging results prior to subsequent statistical and algorithmic information theoretic analyses. Amat *et al.* [24] state that extending our previous work on Lever 2-D to "3D+t is not straight forward." Their survey, as well as our own regarding visualization and editing tools, was unable to find packages that either had an efficient 3-D viewer or one that could handle cell lineages. Lever 3-D fills both of these needs; most notably by creating an efficient 3-D viewer.

Visualization comparison between 2-D projection and stereoscopic 3-D rendering is difficult to quantify. Confocal microscopes are able to capture true three-dimensional data and there are many tools that make two-dimensional projections of this data, such as that of Schmid *et al.* [25]. However, 2-D projection relies on visual cues to convey the relative depth between objects as explained by Wan *et al.*[17]. Using a stereoscopic projection utilizes our binocular vision to convey this information. Even without depth peeling or lighting, the user can perceive depth between object; although, both depth peeling and lighting make the scene look more natural and lowers fatigue. With this perceived depth, validation can be more accurate and efficient. Stereoscopic projection can also help earlier in the processing pipeline by expediting discovery. Interactions between structures in the SVZ are not fully known. Direct stereoscopic observation facilitates the identification of regions to quantify and determine their significance. This discovery phase optimizes the processing pipeline by identifying more precisely what models the processing phase can emulate. We believe that there is enough qualitative benefit to the stereoscopic projection to base a large part of visualization decisions upon it.

## Conclusions

We have developed a new application called LEVER 3-D for validating and correcting the automated segmentation, tracking and lineaging of stem cells from 5-D time lapse image sequence data. The segmentation and tracking results are overlaid on the image data in the 3-D rendering window. The lineage tree for the currently selected clone is shown in a MATLAB 2-D window. Navigation and editing can take place on either window; the MATLAB Mex interface is used to communicate between the C++, CUDA, DirectX, and MATLAB components. The ability to visualize the image data simultaneously with segmentation, tracking, and lineaging makes it possible to quickly identify and easily correct any errors in the automatic analysis. Direct 3D is used for 3-D rendering, providing active shutter stereoscopic visualization and interactive rendering on low-cost gaming hardware. We use the open-source Bioformats tool to read the image data directly from the microscope and CUDA kernels to implement the background removal and segmentation algorithms. The open-source MAT tracking algorithm developed previously for 2-D stem cell image sequences has been enhanced to work with 5-D stem cell data.

One drawback to our use of Direct 3D to enable active shutter stereoscopic is that our application is only available on the Windows operating system. The use of OpenGL would have required explicit application support for active shutter stereoscopic visualization and also the use of the more expensive Quadro-class video cards together with additional RAM for quad buffering, but would have allowed for true cross-platform support. Another drawback of Direct 3D is the lack of a web integration module such as WebGL, which makes it difficult to implement a web client for demonstrating the capabilities of the system or for implementing distributed applications for validating and correcting the 5-D image sequence data. We believe that these shortcomings are offset by the improved visualization available for low-cost from DirectX, with active shutter stereoscopic visualization automatically in the display driver using Nvidia GTX class display cards.

Our goal is to develop an open source solution that allows biologists to process, validate and analyze 5-D stem cell image sequence data in the laboratory, increasing the pace of discovery by combining accurate unsupervised image analysis together with intuitive visualization and validation tools. Data collection has begun for a number of biological experiments that will utilize LEVER 3-D in a high throughput capacity to quantify dynamic behaviors and niche associations for clones of NSCs. The application described here represents a first step in disseminating widely applicable software tools for the analysis of proliferating cells and vasculature from 5-D image sequence data.

**List of abbreviations used**
NSC: *Neural stem cells*, AITPD: *Algorithmic Information Theoretic Prediction and Discovery*, SVZ: *subventricular zone*, SNR: *signal-to-noise ratio*, MAT: *Multitemporal Association Tracking*.

**Competing interests**
The authors declare that they have no competing interests.

**Authors' contributions**
EW implemented algorithms, designed the user interface, and integrated the original LEVER program into the new imaging paradigm. MW was instrumental in the design of the shader paradigm, designing the MEX interface, and provided underling theory to the registration algorithm. CB and YW prepared the tissue samples and captured the subsequent images of both the time lapse and the montage. EK and YW prepared the tissue samples for the time lapse, SG captured the subsequent timelapse movies. ST principle investigator and provided oversight in the biological laboratory. EW and AC wrote the paper. All authors have read and approved the final manuscript.

**Acknowledgments**
Portions of this research were supported by Drexel University, by grant number R01NS076709 from the National Institute Of Neurological Disorders and Stroke, and by the National Institute On Aging of the National Institutes of Health under award number R01AG041861. The content is solely the responsibility of the authors and does not necessarily represent the official views of Drexel or the National Institutes of Health.

**Author details**
[1]Drexel University, 19104 Philadelphia, US. [2]Neural Stem Cell Institute, 12144 Rensselaer, US. [3]UT Health Science Center, 78229 San Antonio, US.

### References

1. Shen, Q., Wang, Y., Kokovay, E., Lin, G., Chuang, S.M., Goderie, S.K., Roysam, B., Temple, S.: Adult svz stem cells lie in a vascular niche: a quantitative analysis of niche cell-cell interactions. Cell Stem Cell **3**(3), 289–300 (2008). doi:10.1016/j.stem.2008.07.026
2. Kokovay, E., Goderie, S., Wang, Y., Lotz, S., Lin, G., Sun, Y., Roysam, B., Shen, Q., Temple, S.: Adult svz lineage cells home to and leave the vascular niche via differential responses to sdf1/cxcr4 signaling. Cell Stem Cell **7**(2), 163–73 (2010). doi:10.1016/j.stem.2010.05.019
3. Tavazoie, M., Van der Veken, L., Silva-Vargas, V., Louissaint, M., Colonna, L., Zaidi, B., Garcia-Verdugo, J.M., Doetsch, F.: A specialized vascular niche for adult neural stem cells. Cell Stem Cell **3**(3), 279–88 (2008). doi:10.1016/j.stem.2008.07.025
4. Cohen, A.R., Bjornsson, C., Temple, S., Banker, G., Roysam, B.: Automatic summarization of changes in biological image sequences using algorithmic information theory. IEEE Trans Pattern Anal Mach Intell **31**(8), 1386–1403 (2009)
5. Cohen, A.R., Gomes, F., Roysam, B., Cayouette, M.: Computational prediction of neural progenitor cell fates. Nat Methods **7**(3), 213–218 (2010)
6. Winter, M., Wait, E., Roysam, B., Goderie, S., Kokovay, E., Temple, S., Cohen, A.R.: Vertebrate neural stem cell segmentation, tracking and lineaging with validation and editing. Nature Protocols **6**(12), 1942–52 (2011). doi:10.1038/nprot.2011.422
7. Doetsch, F., Caille, I., Lim, D.A., Garcia-Verdugo, J.M., Alvarez-Buylla, A.: Subventricular zone astrocytes are neural stem cells in the adult mammalian brain. Cell **97**(6), 703–16 (1999). doi:S0092-8674(00)80783-7 [pii]
8. Ortega, F., Costa, M.R., Simon-Ebert, T., Schroeder, T., Gotz, M., Berninger, B.: Using an adherent cell culture of the mouse subependymal zone to study the behavior of adult neural stem cells on a single-cell level. Nat Protoc **6**(12), 1847–59 (2011). doi:10.1038/nprot.2011.404 nprot.2011.404 [pii]
9. BioFormats. http://loci.wisc.edu/software/bio-formats
10. Michel, R., Steinmeyer, R., Falk, M., Harms, G.S.: A new detection algorithm for image analysis of single, fluorescence-labeled proteins in living cells. Microsc Res Tech **70**(9), 763–70 (2007). doi:10.1002/jemt.20485
11. Otsu, N.: A threshold selection method from gray-level histograms. Systems, Man and Cybernetics, IEEE Transactions on **9**(1), 62–66 (1979)
12. Ceccarelli, M.: A finite markov random field approach to fast edge-preserving image recovery. Image and Vision Computing **25**(6), 792–804 (2007). doi:10.1016/j.imavis.2006.05.021
13. Narayanaswamy, A., Dwarakapuram, S., Bjornsson, C.S., Cutler, B.M., Shain, W., Roysam, B.: Robust adaptive 3-d segmentation of vessel laminae from fluorescence confocal microscope images and parallel gpu implementation. IEEE Trans Med Imaging **29**(3), 583–97 (2010). doi:10.1109/TMI.2009.2022086
14. Al-Kofahi, O., Radke, R.J., Goderie, S.K., Shen, Q., Temple, S., Roysam, B.: Automated cell lineage tracing: a high-throughput method to analyze cell proliferative behavior developed using mouse neural stem cells. Cell Cycle **5**(3), 327–335 (2006)
15. QHULL. http://www.qhull.org/
16. Winter, M., Fang, C., Banker, G., Roysam, B., Cohen, A.: Axonal transport analysis using multitemporal association tracking. International Journal of Computational Biology and Drug Design **(In press)** (2012)
17. Wan, Y., Otsuna, H., Chien, C.B., Hansen, C.: An interactive visualization tool for multi-channel confocal microscopy data in neurobiology research. IEEE Trans Vis Comput Graph **15**(6), 1489–96 (2009). doi:10.1109/TVCG.2009.118
18. Corporation, N.: Nvidia 3-d vision automatic best practices guide. Report (July 2010). http://developer.download.nvidia.com/whitepapers/2010/NVIDIA%203D%20Vision%20Automatic.pdf
19. Murray, J.I., Bao, Z., Boyle, T.J., Waterston, R.H.: The lineaging of fluorescently-labeled caenorhabditis elegans embryos with starrynite and acetree. Nat Protoc **1**(3), 1468–76 (2006). doi:10.1038/nprot.2006.222
20. Aydin, Z., Murray, J.I., Waterston, R.H., Noble, W.S.: Using machine learning to speed up manual image annotation: application to a 3d imaging protocol for measuring single cell gene expression in the developing c. elegans embryo. BMC Bioinformatics **11**(1) (2010). doi:10.1186/1471-2105-11-84
21. Walter, T., Shattuck, D.W., Baldock, R., Bastin, M.E., Carpenter, A.E., Duce, S., Ellenberg, J., Fraser, A., Hamilton, N., Pieper, S., Ragan, M.A., Schneider, J.E., Tomancak, P., Heriche, J.K.: Visualization of image data from cells to organisms. Nat Methods **7**(3 Suppl), 26–41 (2010). doi:10.1038/nmeth.1431
22. Schroeder, W., Martin, K., Lorensen, B.: The Visualization Toolkit, 2nd edn., p. 645. Prentice Hall PTR, Upper Saddle River, NJ (1998)
23. de Chaumont, F., Dallongeville, S., Olivo-Marin, J.C.: Icy: A new open-source community image processing software. In: Biomedical Imaging: From Nano to Macro, 2011 IEEE International Symposium On, pp. 234–237
24. Amat, F., Keller, P.J.: Towards comprehensive cell lineage reconstructions in complex organisms using light-sheet microscopy. Development, Growth and Differentiation (2013). doi:10.1111/dgd.12063
25. Schmid, B., Schindelin, J., Cardona, A., Longair, M., Heisenberg, M.: A high-level 3d visualization api for java and imagej. BMC Bioinformatics **11**, 274 (2010). doi:10.1186/1471-2105-11-274. Schmid, Benjamin Schindelin, Johannes Cardona, Albert Longair, Mark Heisenberg, Martin England BMC Bioinformatics. 2010 May 21;11:274.